\documentclass[%
twocolumn,%
aip,%
reprint,%
]{revtex4-1}

\usepackage[english]{babel}
\usepackage[utf8]{inputenc}
\usepackage[caption=false]{subfig}
\usepackage{graphicx}
\usepackage{dcolumn} 
\usepackage{bm}      
\usepackage{color}   
\usepackage{latexsym}
\usepackage{amsmath}
\usepackage{amssymb}
\usepackage{amsfonts}
\usepackage[space-before-unit,range-units = repeat]{siunitx}

\newcommand {\txt}[1]{\text{#1}}
\newcommand {\brc}[1]{\left( #1 \right)}

\DeclareSIUnit\atmosphere{atm}

\graphicspath{{Figures/}}

\begin{document}

%\preprint{AIP/123-QED}

\title{Drop splashing is independent of substrate wetting}

\author{Andrzej Latka}
\email{alatka@uchicago.edu}
\affiliation{The James Franck Institute and Department of Physics,
The University of Chicago, 929 E 57th Street, Chicago, Illinois 60637, USA}

\author{Arnout M. P. Boelens}
\affiliation{Institute for Molecular Engineering, The University of Chicago, 5801
South Ellis Avenue, Chicago, Illinois 60637, USA}

\author{Sidney R. Nagel}
\affiliation{The James Franck Institute and Department of Physics,
The University of Chicago, 929 E 57th Street, Chicago, Illinois 60637, USA}

\author{Juan J. de Pablo}
\affiliation{Institute for Molecular Engineering, The University of Chicago, 5801
South Ellis Avenue, Chicago, Illinois 60637, USA}

\date{\today}

\begin{abstract}
A liquid drop impacting a dry solid surface with sufficient kinetic energy will
splash, breaking apart into numerous secondary droplets. This phenomenon shows
many similarities to forced wetting, including the entrainment of air at the
contact line. Because of these similarities and the fact that forced wetting has
been shown to depend on the wetting properties of the surface, existing theories
predict splashing to depend on wetting properties as well. However, using
high-speed interference imaging we observe that at high capillary numbers
wetting properties have no effect on splashing for various liquid-surface
combinations. Additionally, by fully resolving the Navier-Stokes equations at
length and time scales inaccessible to experiments, we find that the shape and
motion of the air-liquid interface at the contact line at the edge of the
droplet are independent of wettability. We use these findings to evaluate
existing theories and to compare splashing with forced wetting.
\end{abstract}

%\pacs{47.20Gv, 47.55.dr, 47.55.np.}

%\keywords{}

\maketitle

%===================================================
\section{Introduction}
%===================================================
At first glance, one might suppose that the same physics should describe a solid plunging into a liquid
and a liquid drop impacting a solid: both scenarios revolve around a
liquid-gas-solid contact line that is forced to move at large velocities. In the
former case, exceeding a critical contact line velocity leads to the
destabilization of the contact line and the entrainment of gas bubbles in the
liquid. This phenomenon, called dewetting \cite{marchand2012} or wetting failure
\cite{vandre2012}, is also observed in drop impact \cite{driscoll2011}. It has
therefore been suggested that the onset of contact line instability can serve as
an onset criterion for drop splashing \cite{rein2008,riboux2014}. 
However, contrary to intuition, in this work we find that splashing is independent of the
wetting properties of the surface \cite{goede2017}. This result helps us to
further our understanding of the many processes that rely on splashing droplets,
including erosion, coating, cleaning, cooling, high-throughput drug screening, and
different printing technologies \cite{josserand2016,price2014}.

The wetting of a solid by a liquid depends on many parameters, including
viscosity, surface tension, contact line velocity, impurities in the liquid, and
roughness and heterogeneities of the substrate \cite{bonn2009}. In
addition, there is the complication that for a moving contact line the
classical fluid-mechanics assumption of a no-slip condition on the wall breaks
down \cite{huh1971}, and that due to strong local curvature at a contact line
the observed contact angle is not necessarily the same as the microscopic
contact angle \cite{dussan1979}. For real substrates, which show contact angle
hysteresis due to surfaces roughness and chemical heterogeneity, the situation is even
more complicated; while it is known from experiments that surface roughness can either enhance or
reduce splashing depending on its characteristic length scale \cite{latka2012},
in general its effect on wetting and contact angle hysteresis is poorly
understood \cite{bonn2009}.

For smooth surfaces experiments on forced wetting have typically focused on the
relationship between the velocity of the edge of the liquid/gas interface and
the observed dynamic contact angle. Typically, in the steady state case when the
contact angle is plotted as a function of the non-dimensionalized edge velocity,
i.e. the capillary number, a single curve is found
\cite{hoffman1975,ralston2008}. For low capillary numbers and a completely
wetting surface it has been experimentally well established \cite{berg1993} that
the contact line moves according to the Tanner-Voinov law
\citep{voinov1976,tanner1979}. For partly wetting liquids with a sufficiently
high viscosity the data can be described by a variation of the same law without
the assumption of small slopes \cite{cox1986}. For forced wetting at larger
capillary numbers the contact line eventually becomes unstable, and this is a
topic of active research \cite{marchand2012}.

Models of forced wetting \cite{snoeijer2007,marchand2012} use the wetting properties of the
surface as a boundary condition to determine the stability threshold for the
contact line. Indeed, the wettability of the object that is plunged into a liquid has been found to have a strong
influence on wetting failure \cite{duez2007}. In contrast to recent conclusions
based on simulations \cite{yokoi2011,josserand2016}, we present
experimental results which show that for rapidly moving contact lines, for a
wide array of liquids, the surface wettability has negligible effect on splashing. We
also describe simulations that are able to resolve contact line behavior at high
resolution. These simulations reveal that the contact line motion previously
associated with splashing on completely wetting surfaces \cite{boelens2016a}, is
nearly identical for a completely non-wetting surface. Before dewetting, the
rapidly moving contact line in both the wetting and the non-wetting case shows a
microscopic contact angle of $\ang{180}$. This suggests that the assumption of a
fixed microscopic contact angle is inapplicable to contact lines that are forced
to move at high speeds, as is the case in splashing, and challenges theories
based on this assumption.

%===================================================
\section{Methods}
%===================================================
\subsection{Experiments}
%===================================================
\begin{figure}
\includegraphics[width=\columnwidth]{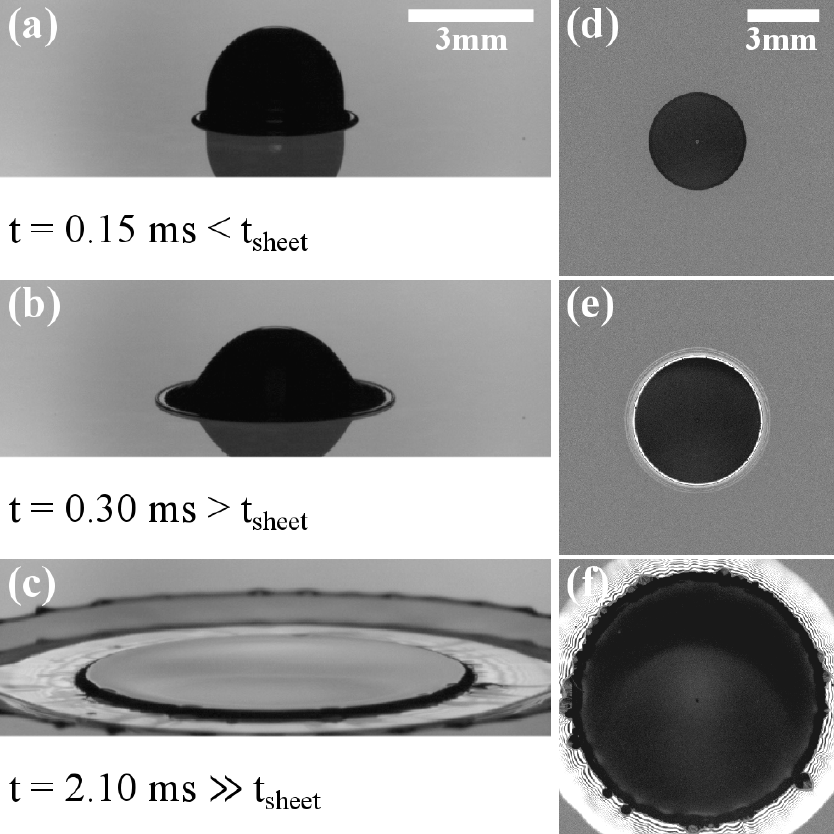}
\caption[]{\label{fig:images} Successive images of a $9.4 \si{\centi P}$
silicone oil drop of diameter 3.2 mm impacting a glass slide at 3.3 m/s at
atmospheric pressure. Images (a) - (c) show the side view of the drop and its
reflection, while (d) - (f) show the interference measurement from below at
corresponding times. Initially, the drop spreads fully wetting the substrate in
the form of a lamella. After time $t_\txt{sheet}=0.21 \si{\m \s}$ an air gap
appears between the liquid and solid, seen clearly as bright interference
fringes (e,f), resulting in the creation of a thin liquid sheet (b,c) that
extends from the thicker lamella.}
\end{figure}    

\begin{figure}
\includegraphics[width=\columnwidth]{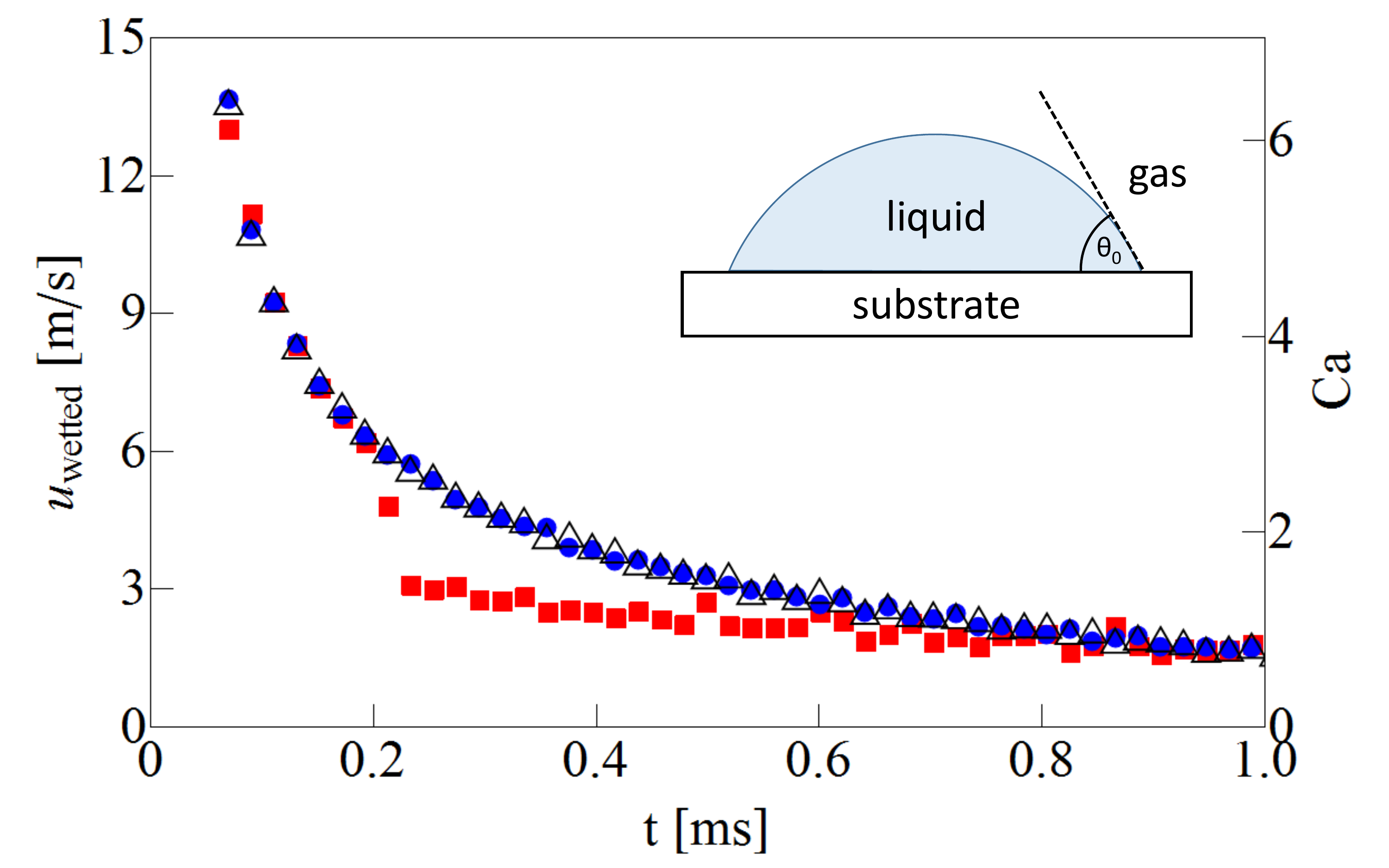}
\caption[]{Speed (left axis) and capillary number (right axis) of the advancing liquid-solid contact line as a
function of time after impact of a $9.4 \si{\centi P}$ silicone oil drop on
glass ($\theta_{0} = \ang{0}$,
\includegraphics[height=6pt]{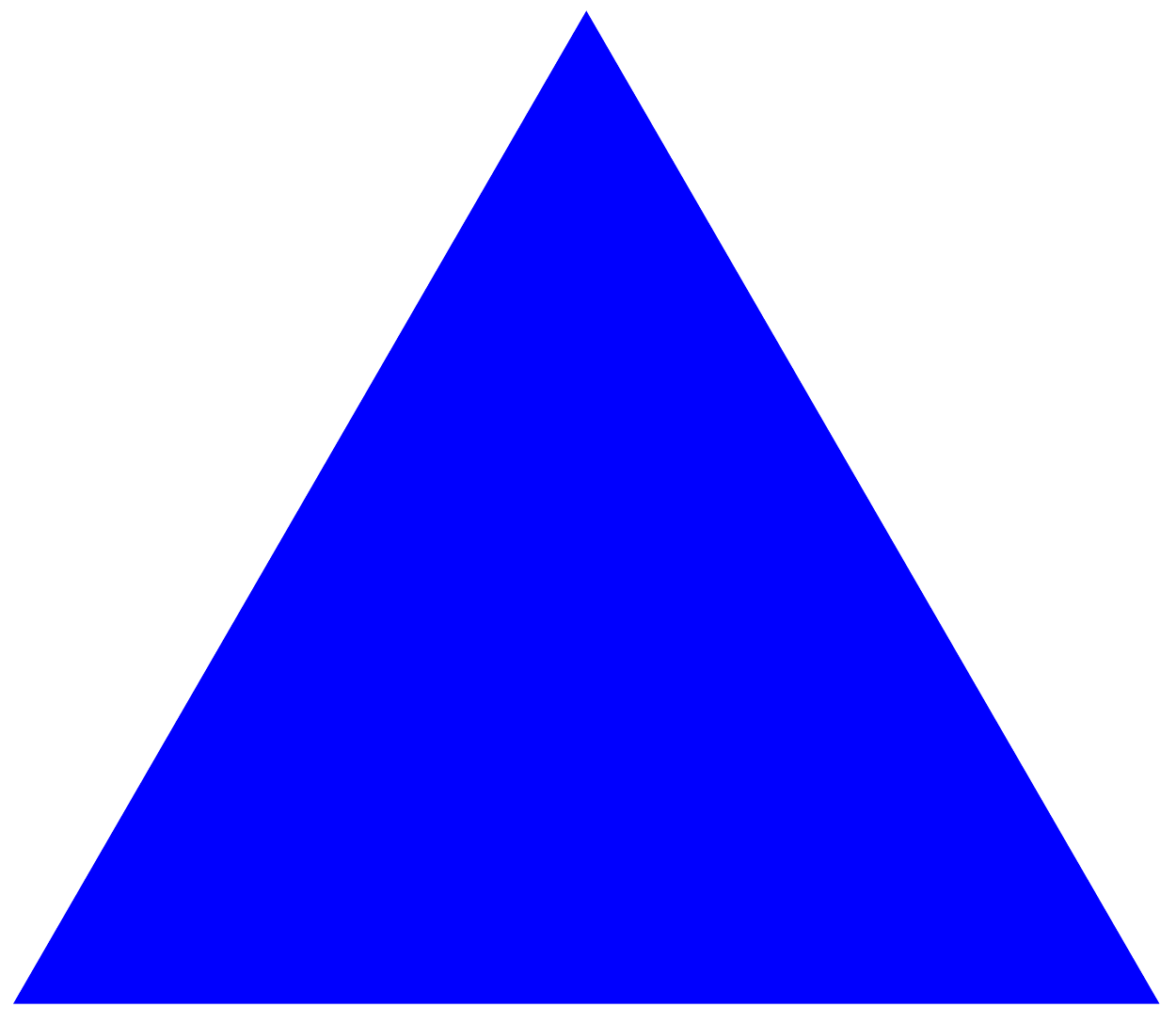}) or glass coated with
an oleophobic layer ($\theta_{0} = \ang{42}$,
\includegraphics[height=6pt]{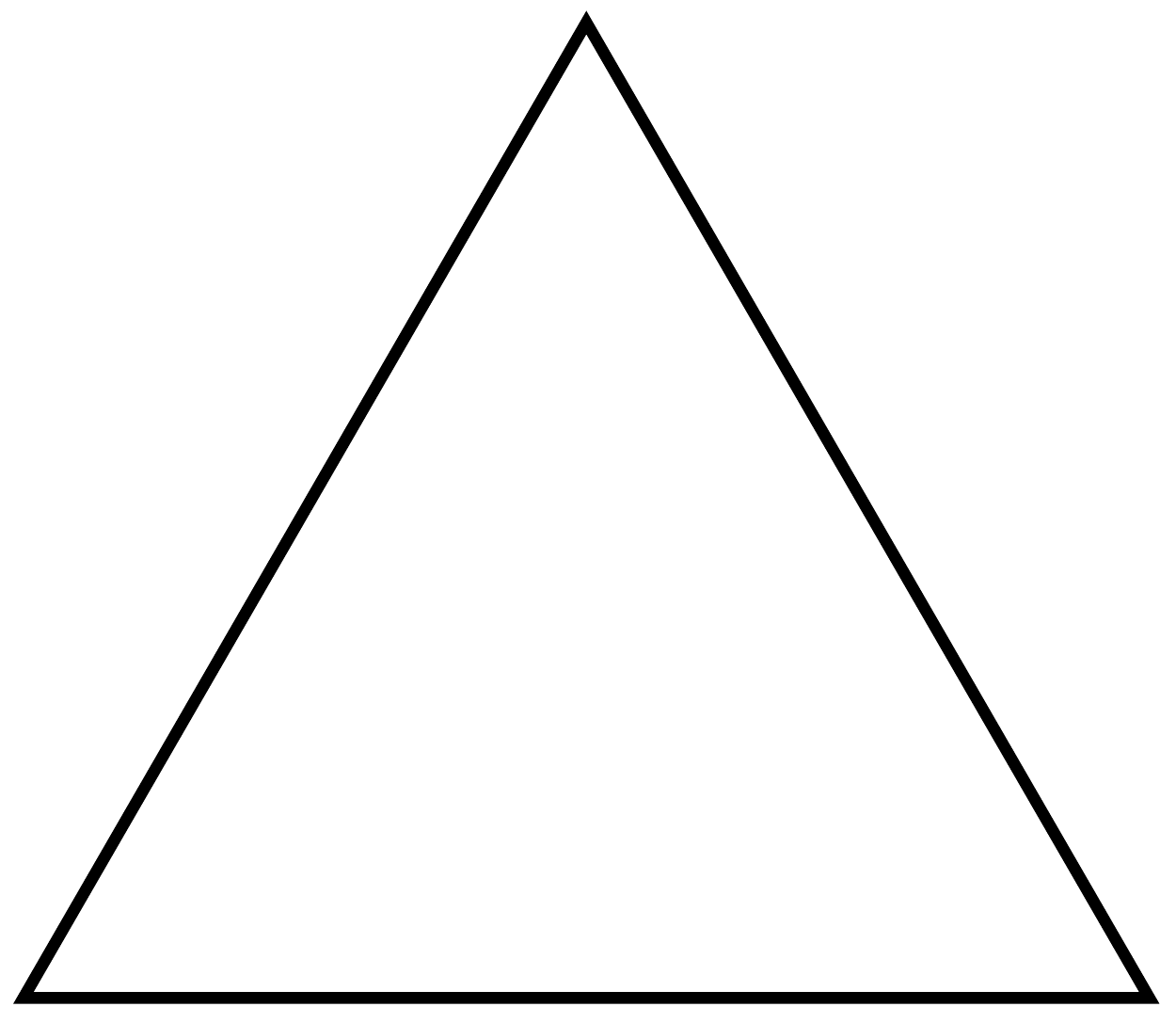}). Splashing was
suppressed by reducing the pressure to $P=30 \si{\kilo Pa}<P_{\txt{sheet}}$. The
presence of a coating does not influence spreading speed. At atmospheric
pressure (\includegraphics[height=6pt]{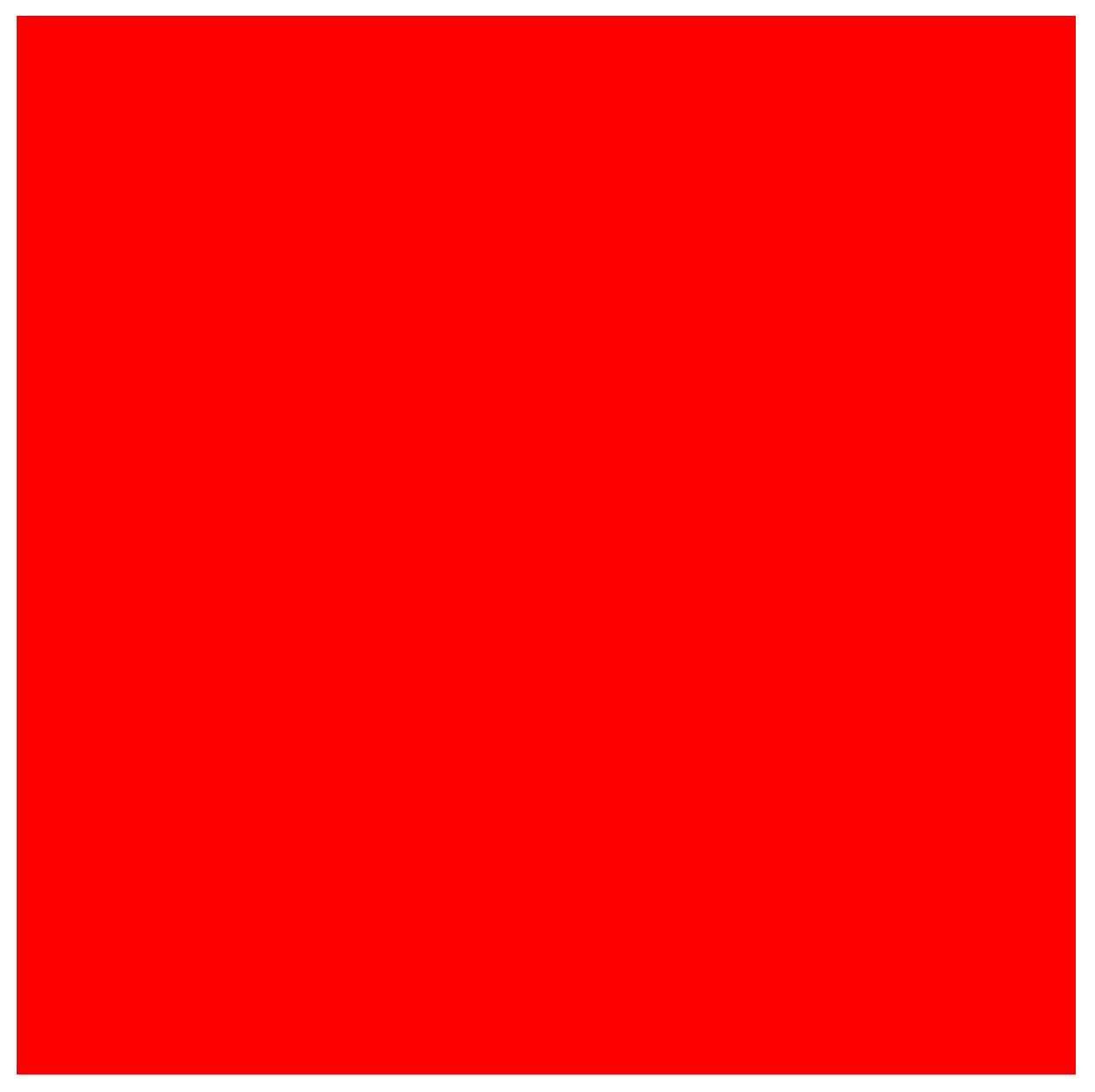}) air entrainment
begins at $t_{\txt{sheet}}=0.21 \si{\milli \s}$ after which the contact line
slows. The diagram shows that Young's angle $\theta_0$, at which a stationary
gas-liquid interface meets the substrate, is measured in the liquid phase.}
\label{fig:uL} 
\end{figure}

The experiments were conducted with either silicone oil (PDMS, Clearco Products)
of viscosity $\mu=9.4$ and $32 \si{\centi P}$ or a mixture of water and glycerol
($\mu=32\si{\centi P}$). The drops with diameter $D=3.3\pm 0.1 \si{\milli \m}$
were produced at a nozzle with a syringe pump and were accelerated by gravity to
an impact velocity of $V = 3.4 \pm 0.1 \si{\m.\s^{-1}}$. This resulted in the
non-dimensional numbers and ratios shown in Tab.~\ref{tab:nondim}. 
The silicone oil drops
impacted glass slides (Fisherbrand Microscope Slides) that were left untreated
to provide a completely wetting surface, with contact angle $\theta_{0} = \ang{0}$, or
glass slides covered with an oleophobic coating (Fussode Coat, $\theta_{0} = 42
\pm \ang{2}$). Similarly, the water-glycerol drops impacted either clean glass (
$\theta_{0} = 36 \pm \ang{3}$), glass coated with indium tin oxide ( $\theta_{0}
= 79 \pm \ang{4}$) or a hydrophobic coating (RainX, $\theta_{0} = 90 \pm
\ang{3}$). A wetting substrate was achieved for the water-glycerol by
pre-wetting the glass slide with the same mixture. A liquid drop fully wets such
a prepared slide ( $\theta_{0} = \ang{0}$), however the coating is thin enough
not to change the splashing behavior. The changing of substrates only affects
the contact angle and does not change the spreading dynamics.
Fig.~\ref{fig:uL} shows that drops spread at the same rate regardless of
$\theta_0$. Air was used as the ambient gas, whose pressure $P$ was controlled
in a vacuum chamber $(5 \si{\kilo Pa} \le P \le 101 \si{\kilo Pa})$. Impacts
were recorded with high-speed cameras (Vision Research) at rates up to 130 000
fps either from the side as in Fig.~\ref{fig:images}(a-c), or with interference
imaging (d-f). The latter method measures the interference between light
reflected from the bottom surface of the spreading liquid and the top surface of
the substrate. Wherever the liquid is in contact with the substrate, there is no
reflection of light from that surface and thus no light entering the camera.
Wherever the two are separated, an interference pattern is created, as seen in
Fig.~\ref{fig:images}(e-f). Since this method is particularly sensitive to the
presence of the air gap, it allows us to measure precisely when the contact line
begins to entrain air \cite{driscoll2011}.

A typical splash is presented in the left column of Fig.~\ref{fig:images}.
Fig.~\ref{fig:images}(a) shows that a drop does not splash immediately.
Instead the liquid spreads radially outward in the form of a lamella
\cite{driscoll2010,stevens2014b}. The simultaneous interference image shows that
the lamella fully wets the substrate. Beginning at time $t_{\txt{sheet}}$ after
impact (where we define $t_{sheet}$ as the time when the thin sheet first starts
to appear at the front of the expanding lamella) one can observe an interference
pattern at the liquid-air-solid contact
line, as in Fig.~\ref{fig:images}(e), indicating the presence of a gas film
that is of order a micron thick that separates the spreading liquid from the
substrate. The time $t_\txt{sheet}$ is the start of the formation of a thin
sheet of liquid \cite{driscoll2010, driscoll2011}, as can be seen in
Fig.~\ref{fig:images}(b). The thin-sheet grows and ultimately breaks up into the
secondary droplets that form the splash ($t=2.1 \si{\m s}$,
Fig.~\ref{fig:images}(c,f)). 

The thin-sheet creation time depends on a number of parameters
\cite{driscoll2010}. Most importantly, $t_{\txt{sheet}}$ is delayed as the
ambient gas pressure is reduced. However, if the pressure is decreased below a
threshold pressure $P_{\txt{sheet}}$, instead of being further delayed, the thin
sheet will fail to appear and the splash will have been completely suppressed
\cite{xu2005,xu2007,xu2007b}. We quantify the effect of wetting on splashing by
measuring the dependence of both $t_{\txt{sheet}}$ and $P_{\txt{sheet}}$ on the
surface properties.

%===================================================
\subsection{Simulations}
%===================================================
To be able to examine the contact line in more detail, we simulate the breakup
of a splashing drop using a finite volume implementation of the volume of fluid
method \cite{hirt1981}. The VOF approach evolves around the
definition of a phase parameter $\alpha$ with the following properties:
\begin{equation}
  \alpha
=
  \left\{
  \begin{array}{ll}
  0      & \txt{in gas phase}    \\
  (0, 1) & \txt{on interface}    \\
  1      & \txt{in liquid phase}
  \end{array}
  \right.
\end{equation}
The evolution of $\alpha$ is calculated using the following transport equation:
\begin{equation}
  \frac{\partial \alpha}{\partial t}
+ \nabla \cdot \brc{\alpha \vec{v}}
+ \nabla \cdot \brc{\alpha \brc{1 - \alpha} \vec{v}_{lg}}
=
  0,
\end{equation}
where $\vec{v}$ is the phase averaged velocity, and $\vec{v}_{lg}$ is a velocity
field suitable to compress the interface. This equation is equivalent to a
material derivative, but rewritten to minimize numerical diffusion
\cite{rusche2002}.

The phase parameter is used to calculate the phase averaged density, $\rho$,
velocity, $\vec{v}$, and viscosity, $\mu$, which are used in the momentum balance:
\begin{equation}
  \frac{\partial \rho \vec{v}}{\partial t}
+ \nabla \cdot \brc{\rho \vec{v} \otimes \vec{v}}
=
- \nabla p
+ \nabla \cdot \brc{\mu \nabla \vec{v}}
+ \rho \vec{g}
- \vec{f},
\end{equation}
and the continuity equation:
\begin{equation}
  \nabla \cdot \vec{v} 
= 
  0.
\end{equation}
In the above equations $t$ is time, $p$ is pressure, $g$ is gravity, $\vec{f}$ is any body force, like
the surface tension force, and $\otimes$ is the dyadic product. To complete the
VOF model, an expression is needed to calculate the surface tension force
$\vec{f}_{\txt{st}}$, and a model is needed for the contact line. The
surface tension force is calculated using the expression \cite{brackbill1992}:
\begin{equation}
  \vec{f}_{\txt{st}}
=
  \sigma_{\txt{st}} \kappa \nabla{\alpha}
\end{equation}
where $\sigma_{\txt{st}}$ is the surface tension coefficient, and $\kappa$ is
the curvature of the interface. 

The effect of varying the Young's angle $\theta_{0}$ from $\ang{0}$ to
$\ang{180}$ is calculated directly through the generalized Navier boundary
condition at the impact wall \cite{qian2003,gerbeau2009}. With this boundary
condition the dynamic contact angle $\theta$ is allowed to vary freely, but a
restoring line-tension force is applied at the contact line whenever the dynamic
angle deviates from $\theta_{0}$. This restoring force is an additional source
term in the Navier-Stokes equations, and has the following form:
\begin{equation}
  \vec{f}_{\txt{lt}}
=
- \frac{\sigma}{h}
  \cos{\theta_{0}}
  \nabla_{\txt{2D}}{\alpha}
\end{equation}
In the above equation $\sigma$ is the surface tension coefficient,
$h$ is the height of the local grid cell, and $\nabla_{\txt{2D}}{\alpha}$ is the
gradient of $\alpha$ on the wall. This force is applied on the liquid-gas
interface in the first grid cells adjacent to the wall and is balanced by the
surface tension force when $\theta$ is equal to $\theta_{0}$. Away from the
contact line the used implementation of the generalized Navier boundary
condition reduces to the Navier-slip boundary condition. Using this slip 
boundary condition gives a good approximation for the thin film behavior at the
wall \cite{lauga2007,sprittles2017}. Because the model used can accommodate only
one value for the slip length, a value of $\lambda = 1 \si{\nano\m}$ is chosen to
be able to describe the contact line accurately. However, in practice the
effective slip length is on the order of the mesh size of $10 \si{\nano\m}$
\cite{jacqmin2000}. This results in the gas film potentially closing faster in
our simulations than if the slip length were truly $1 \si{\nano\m}$. 

The simulations are performed for ethanol in air using the VOF solver of the OpenFOAM Finite
Volume toolbox \cite{openfoam} at up to $10 \si{\nano \m}$ resolution at the
wall \footnote{
The pressure field is solved using the Preconditioned Conjugate Gradient solver
(PCG) with the Geometric Agglomerated Algebraic Multigrid (GAMG) pre-solver. The
tolerance was set to $10^{8}$ to ensure convergence. For divergence terms a
Gauss Upwind Scheme was used. OpenFOAM regulates the time step size through the
Courant number, which was set to $0.1$ to ensure a small time step and numerical
stability.}.
More information on the boundary conditions at the contact line can be
found in Ref.~\citenum{boelens2016b}. More information on the equations,
initial conditions, and a comparison with experiments van be found in
Ref.~\citenum{boelens2016a}. In this paper it is shown that the scaling of the gas
film height as function of impact velocity is consistent with theory and
experiments, and multiple experimental observations are reproduced, including
the formation of the central air bubble, liquid sheet formation, and contact
line instability \cite{boelens2016a}.

To reduce memory requirements, we consider an ethanol ($\mu = 1.1 \si{\centi
P}$) droplet with a diameter of $300 \si{\micro \m}$, as opposed to the $3
\si{\milli \m}$ more viscous droplets used in the experiments. This results in
the non-dimensional numbers shown in Tab.~\ref{tab:nondim}. As can be seen in
this table the non-dimensional numbers for the whole droplet are quite different
between the experiments and simulations. However, the focus of this work is on
the capillary number at the contact line/edge of the droplet, which, as is shown
below, is of the same order between simulations and experiments. In addition, as
the splashing threshold has been shown to scale across a wide range of
parameters \cite{stevens2014,stevens2014b,riboux2014}, comparisons with
experiments should not be compromised.

\begin{table}[h]
\caption{An overview of the non-dimensional numbers and ratios of liquid and gas
properties under conditions of our experiments.}
\label{tab:nondim}
\begin{tabular}{lrrrr}
                          & $\txt{Re}$ & $\txt{We}$ & $\rho_{l}/\rho_{g}$ & $\nu_{l}/\nu_{g}$ \\
\hline
\hline
Silicone oil (9.4 cP)     & $1106$     & $1759$     & $0.6852$            & $927$             \\
\hline                    
Silicone oil (32 cP)      & $335$      & $1762$     & $2.262$             & $956$             \\
\hline
Water \& glycerol (32 cP) & $418$      & $686$      & $1.825$             & $1191$            \\
\hline
Ethanol                   & $986$      & $264$      & $0.1028$            & $789$             \\
\hline
\hline
\end{tabular}
\end{table}

%===================================================
\section{Results}
%===================================================
\subsection{Experiments}
%===================================================
\begin{figure}
\subfloat{
\includegraphics[width=\columnwidth]{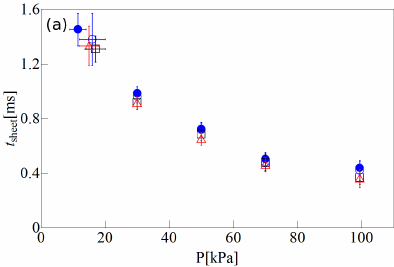}
} \\
\subfloat{
\includegraphics[width=\columnwidth]{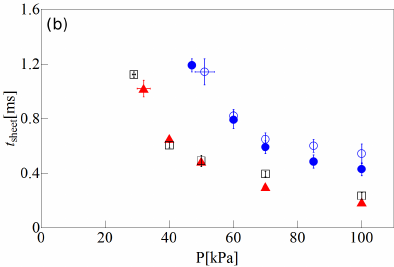}
}
\caption{
Time of thin-sheet creation vs. ambient gas pressure for droplet impacts on glass
slides with different wetting angles. Image (a) shows the impact of $32
\si{\centi P}$ water-glycerol drops on glass slides with $\theta_{0} = \ang{0}$
(\protect\includegraphics[height=6pt]{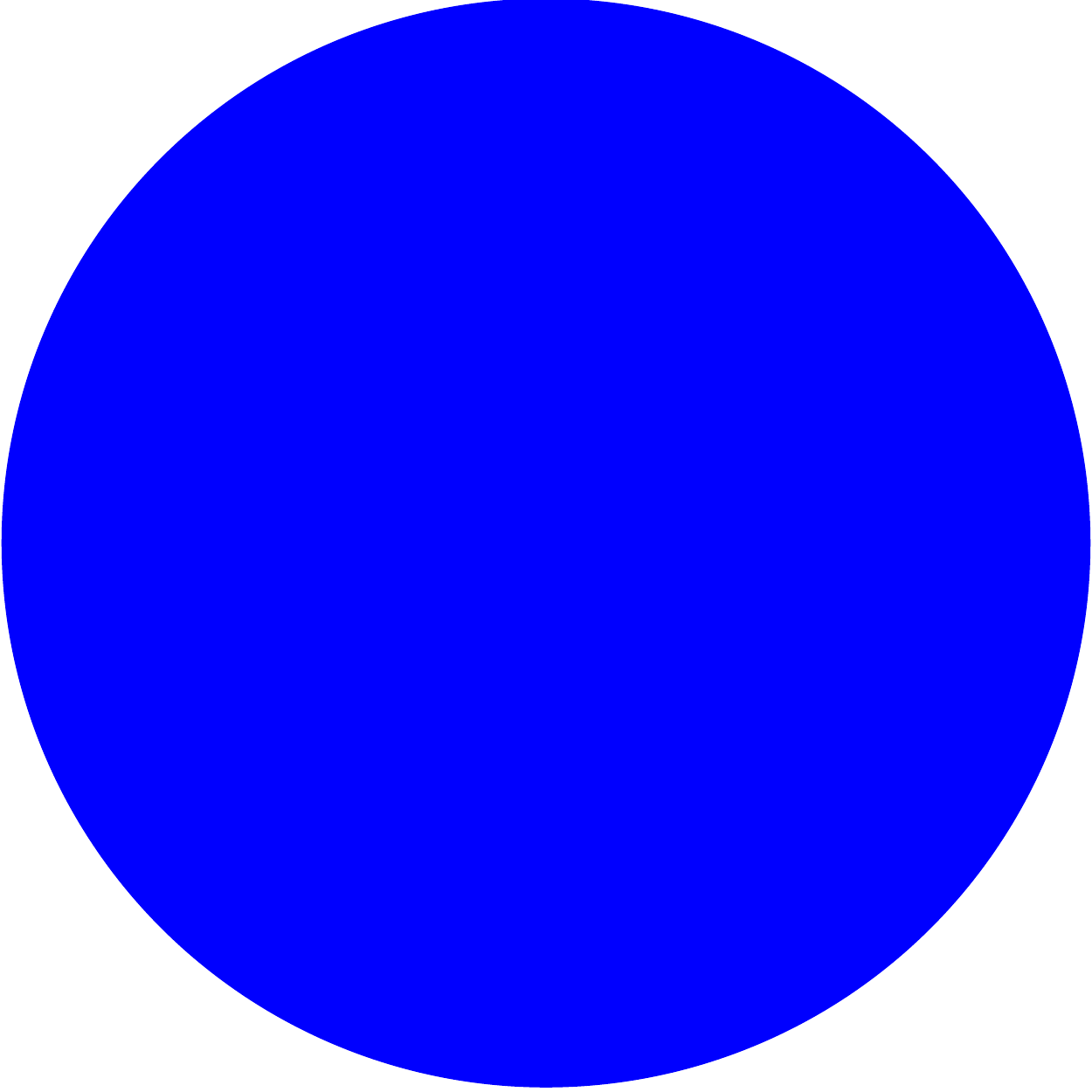}), $\ang{36}$
(\protect\includegraphics[height=6pt]{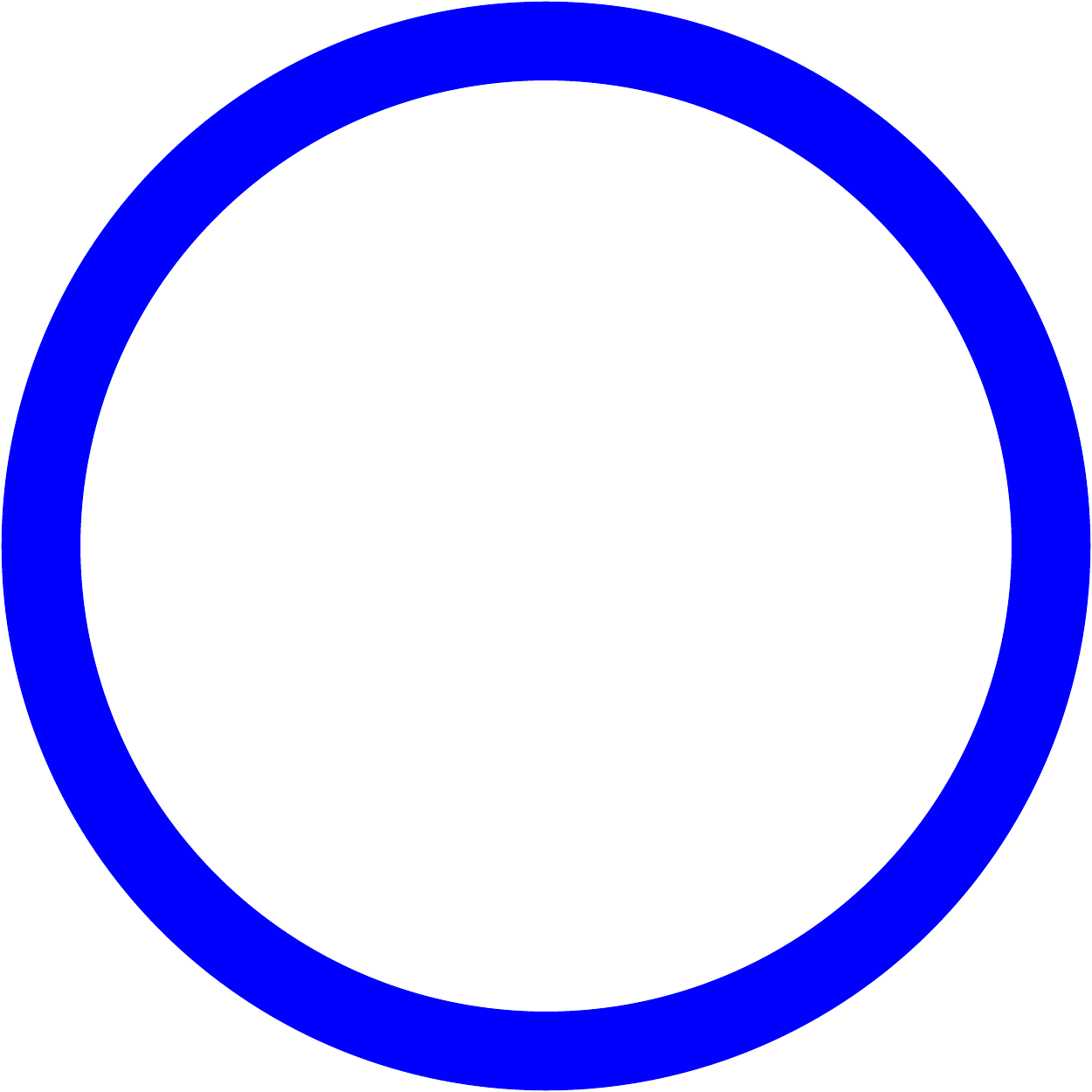}), $\ang{79}$
(\protect\includegraphics[height=6pt]{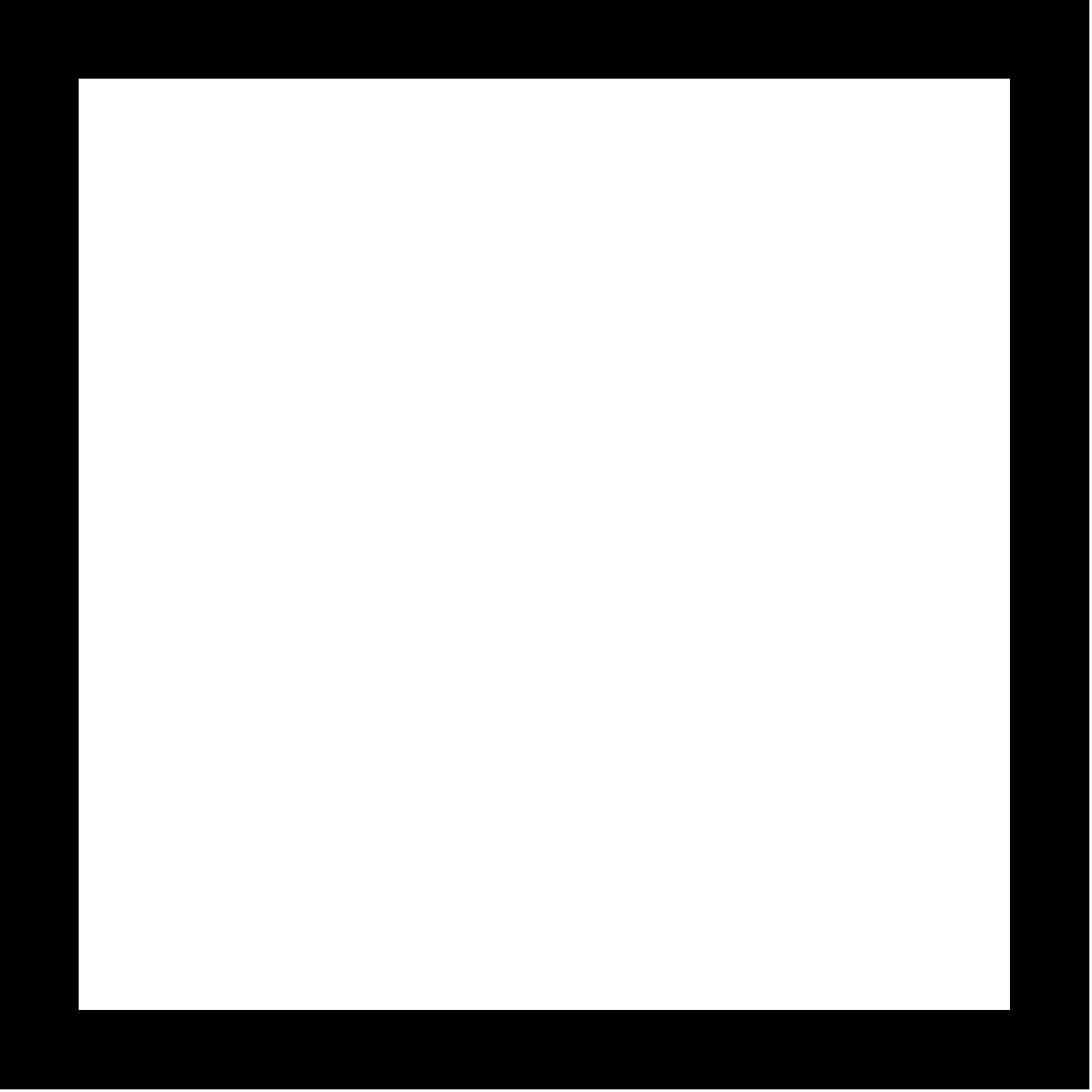}), and $\ang{90}$
(\protect\includegraphics[height=6pt]{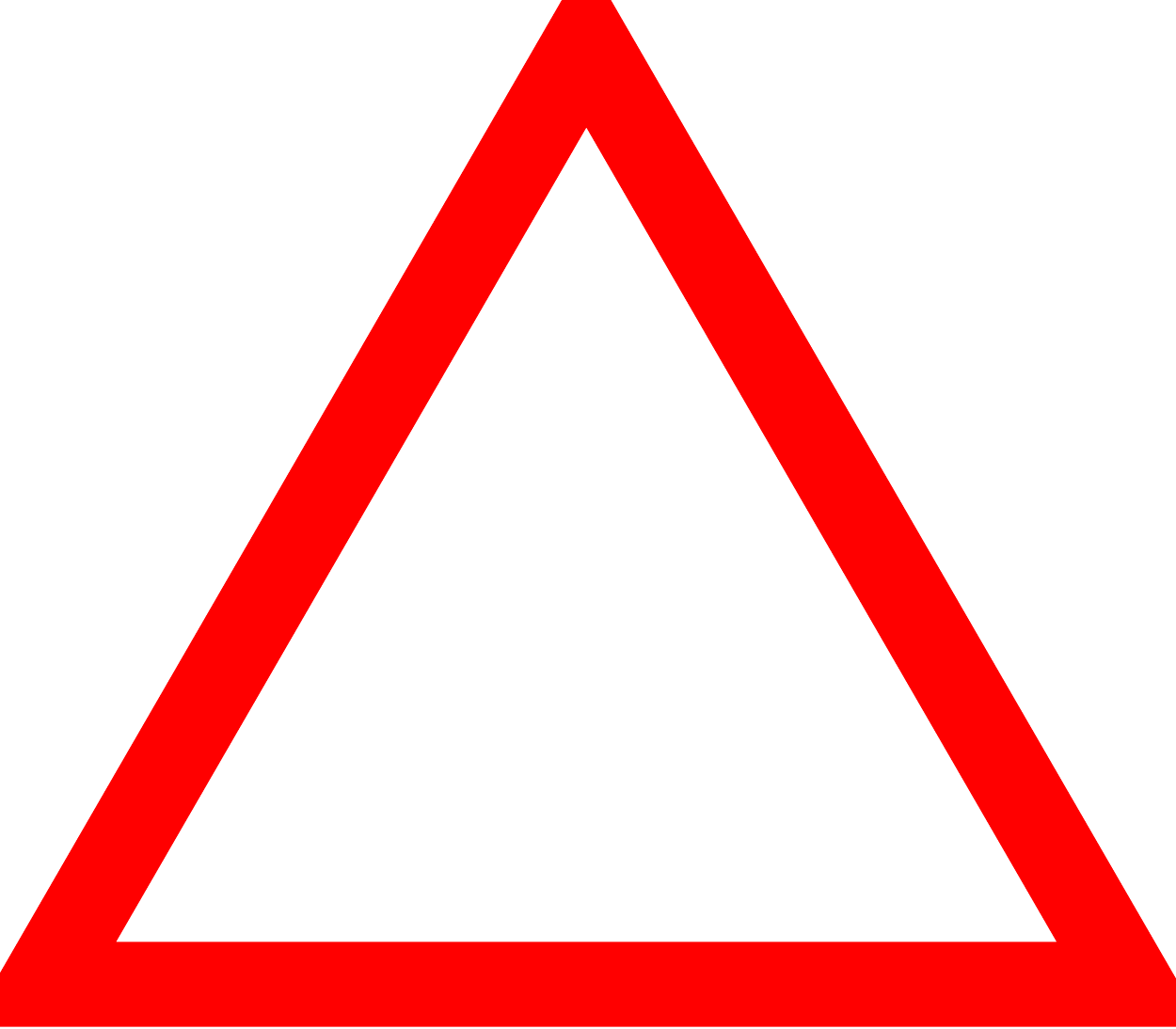}). $t_\txt{sheet}$ is
independent of wetting. Image (b) shows the impact of of $9.4 \si{\centi P}$
silicone oil drops on glass slides with $\theta_{0} = \ang{0}$
(\protect\includegraphics[height=6pt]{Filled_Blue_Circle}) and $\ang{42}$
(\protect\includegraphics[height=6pt]{Empty_Blue_Circle}), and  of $32 \si{\centi
P}$ silicone oil drops on glass slides with $\theta_{0}$ of $\ang{0}$
(\protect\includegraphics[height=6pt]{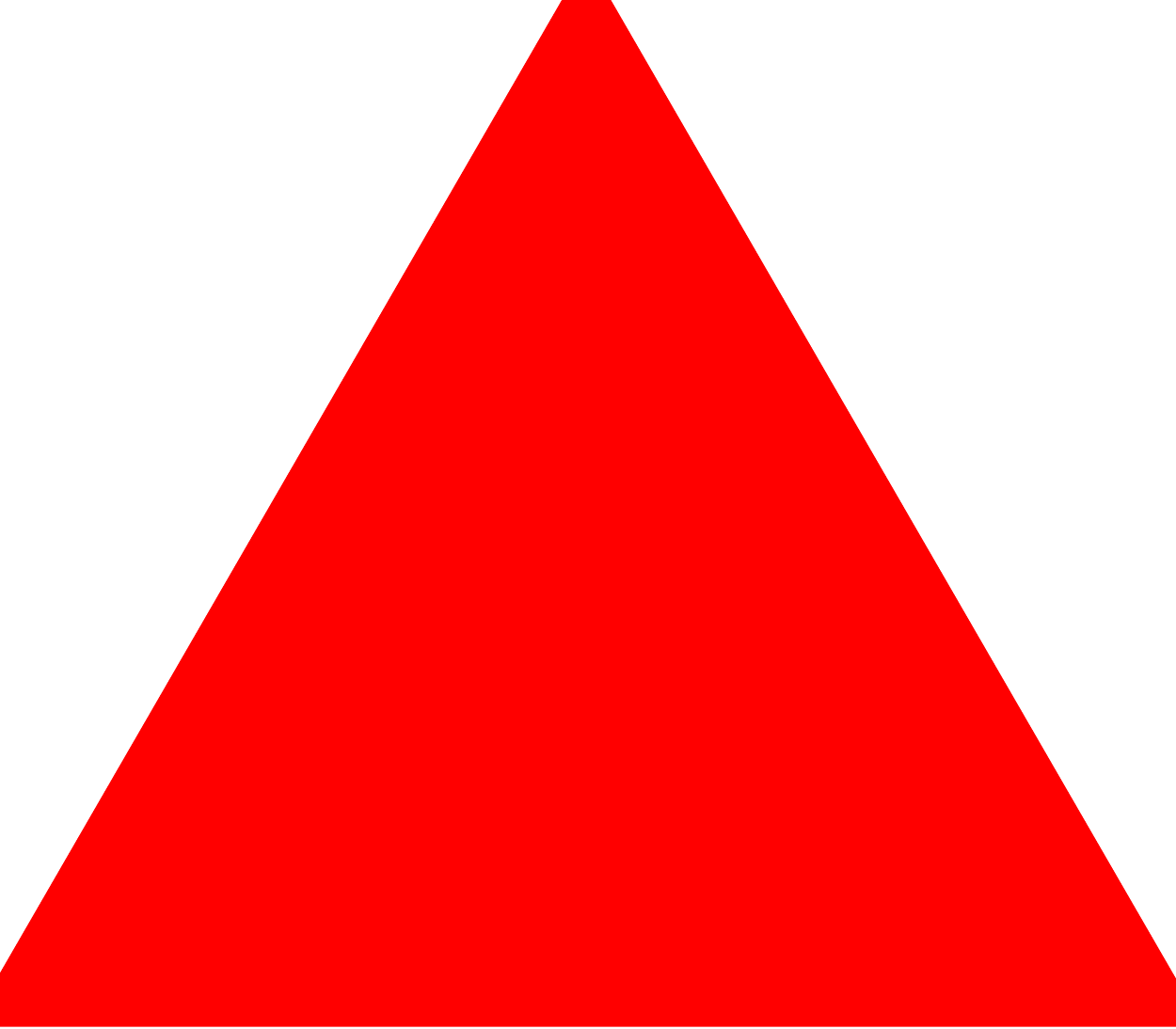}) and $\ang{42}$
(\protect\includegraphics[height=6pt]{Empty_Black_Square}). The small differences in
$t_{\txt{sheet}}$ with wetting properties are within error.
}
\label{fig:tSheet}
\end{figure}

\begin{figure}
\includegraphics[width=\columnwidth]{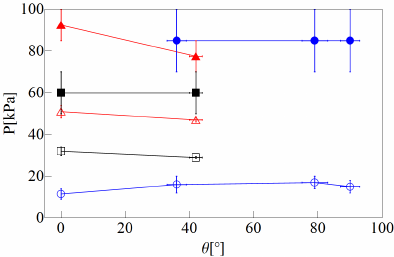}
\caption[]{\label{fig:PT} Threshold pressure for drop breakup $P_{\txt{splash}}$
(filled symbols) and thin-sheet creation $P_{\txt{sheet}}$ (empty symbols) vs.
contact angle: $32 \si{\centi P}$ water-glycerol
(\includegraphics[height=6pt]{Filled_Blue_Circle}), $32 \si{\centi P}$
silicone oil
(\includegraphics[height=6pt]{Filled_Red_UpTriangle}) and $9.4 \si{\centi P}$ silicone oil
(\includegraphics[height=6pt]{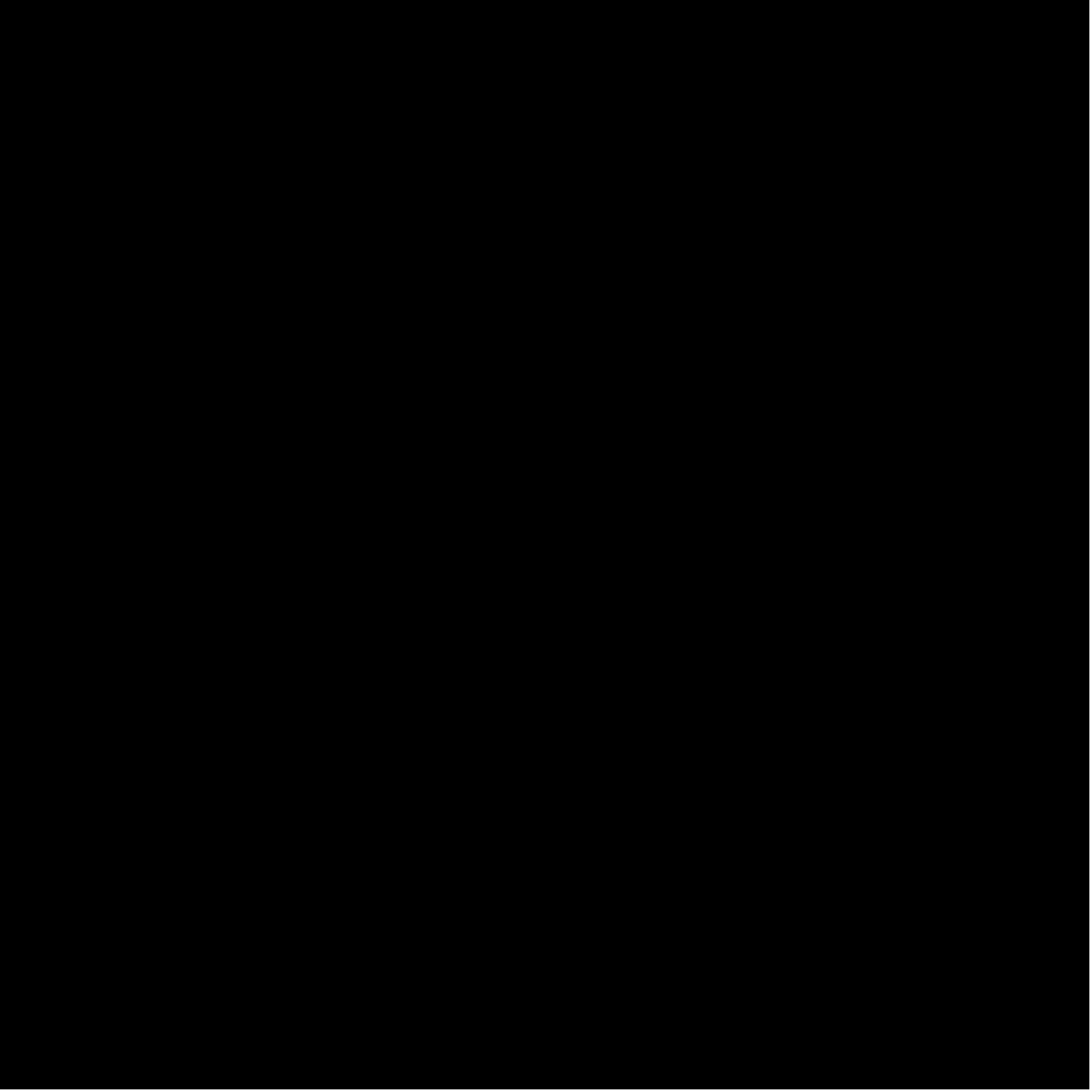}). Thin
sheets created at $P_\txt{sheet}<P<P_\txt{splash}$ will not break apart. Neither threshold is affected by surface wettability.}
\end{figure}

Varying the surface wettability does not affect $t_{\txt{sheet}}$, as shown for
water-glycerol drops in Fig.~\ref{fig:tSheet} (a). Notably, the onset of
thin-sheet creation is surprisingly independent of changes in surface
properties. Not only does a hydrophobic coating fail to change
$t_{\txt{sheet}}$, but even coating the glass slide with a thin layer of
water-glycerol yields the same result. Similarly, no noticeable effect of
wetting can be seen for silicone oil drops in Fig.~\ref{fig:tSheet} (b), where the
substrate was changed from fully wetting with $\theta_{0} = \ang{0}$ to
partially wetting with $\theta_{0} = \ang{42}$. 

Fig.~\ref{fig:PT} compares the threshold pressure for the different
substrates. As the ambient pressure is decreased, $t_{\txt{sheet}}$ is delayed
and the resulting thin-sheet is diminished. Consequently, below a pressure
$P_{\txt{splash}}$, the thin-sheet is too small to break apart into secondary
droplets and splashing is suppressed. If the ambient pressure is further
decreased below $P_{\txt{sheet}}$, the thin-sheet is never formed. We find that
both $P_{\txt{sheet}}$ and $P_{\txt{splash}}$ are independent of $\theta_{0}$.
This result is similar to the velocity threshold in Ref.~\citenum{duez2007}, which
was also independent of $\theta_{0}$ for $\theta_{0} < \ang{90}$.

%===================================================
\subsection{Simulations} \label{sec:ResSim}
%===================================================
\begin{figure*}
\includegraphics[width=\textwidth]{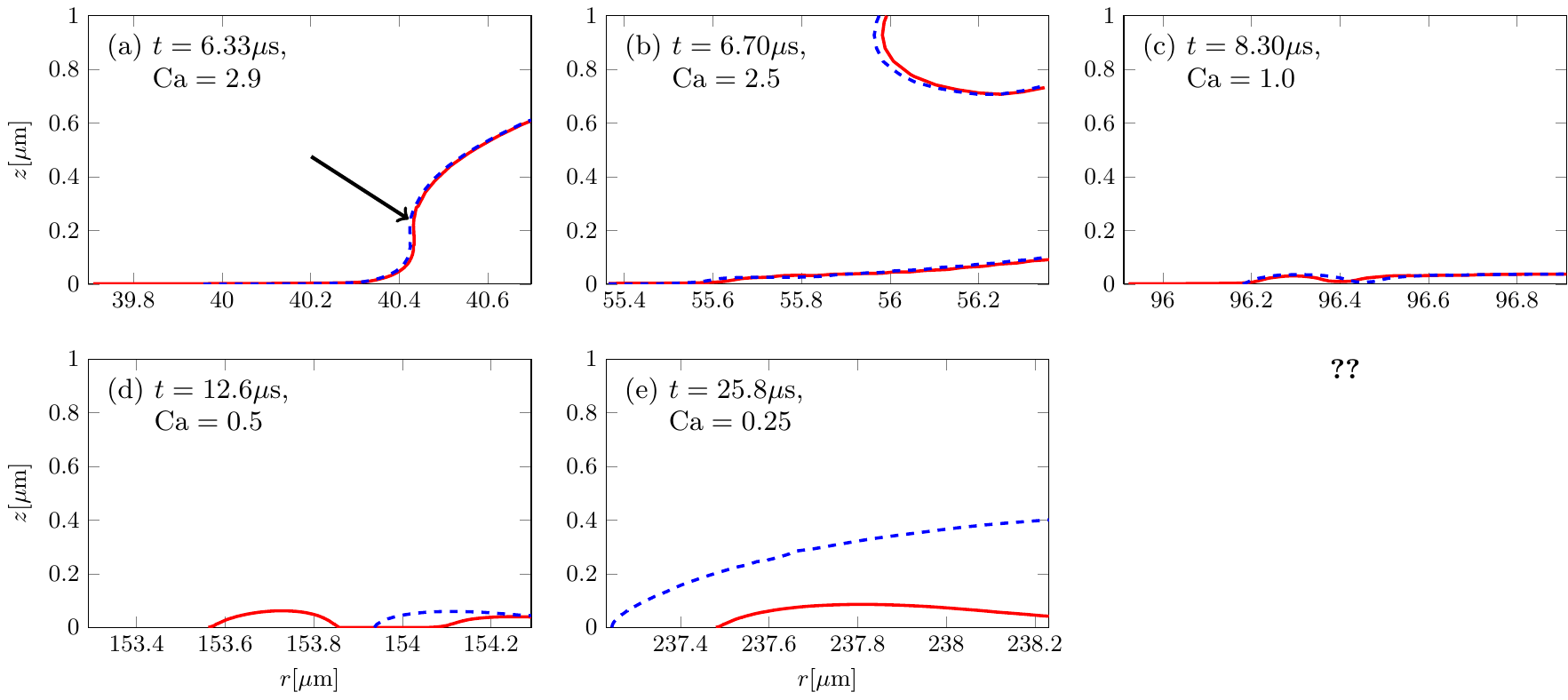} 
\caption{\label{if} Time series of the droplet interface showing the
contact-line evolution for a simulated droplet with the parameters of ethanol at
atmospheric pressure for $\theta_{0} = \ang{0}$ (wetting) and $\theta_{0} =
\ang{180}$ (non-wetting). The vertical axis shows height above the surface and
the horizontal axis shows the radial distance from the point of impact. (a) The
moment a cusp (indicated by the black arrow) can first be observed in the
interface. This is the onset of lamella formation. (b) Immediately after
$t_\txt{sheet}$. A gas film is present under the liquid sheet and the interface
approaches the surface at a $\ang{180}$ angle. (c) The transition from the
$\txt{Ca} > 1$, high-speed contact line regime to the $\txt{Ca} < 1$, low-speed
regime. (d) As explained in section \ref{sec:ResSim}, this frame shows a
touch-down event of the interface on the non-wetting surface: a gas bubble is
entrained behind the contact line. (e) The gas film forming on the wetting
surface at low speeds is thicker than the gas film on the non-wetting surface.}
\end{figure*}

The simulation results shown in Fig.~\ref{if} compare the shape of the
air-liquid interface for wetting and non-wetting surfaces at different stages of
impact. The interface has a thickness of about $20 \si{\nano\m}$, which is two
times the highest resolution of the simulations. To facilitate comparisons with experiments, velocity is
non-dimensionalized as the capillary number $\txt{Ca} \equiv \frac{\mu
V}{\sigma}$. At all times for which $\txt{Ca} \ge 1$ the interface is the same on
both surfaces and, within the resolution of the simulations, the microscopic
contact angle in both cases is $\ang{180}$. The observed gas film is about $20
\si{\nano\m}$ thick.

Only for times at which $\txt{Ca} < 1$ do the interfaces begin to look
different; the contact angle on the non-wetting surface remains $\ang{180}$,
while the contact angle on the wetting surface decreases and would converge to
its equilibrium Young's angle, $\theta_{0} = \ang{0}$, if the simulation were to
run long enough until the drop is stationary. 

A consequence of the lower $\theta$ in the wetting case at low $\txt{Ca}$ is a thicker
gas film at the contact line compared to the non-wetting surface, as can be
observed most clearly in Fig.~\ref{if} (e). Independent of the wetting
properties, the gas film thickness is greatest when the air-liquid interface
becomes parallel to the substrate (cf. Figs.~\ref{if} (c-e)). The lower $\theta$
at the contact line of a wetting substrate requires a longer arc length to
satisfy this condition, which results in a thicker gas film. Additionally, a
greater separation between the liquid sheet and the substrate stabilizes the
contact line. At early times, the gas film in front of the contact line
periodically collapses and the liquid touches down on the substrate, entraining
air bubbles (Fig.~\ref{if} (d)). The contact line on the non-wetting surface
never stabilizes and bubbles are entrained throughout the simulation. In
contrast, these touch-down events are no longer observed at late times on the
wetting surface.

\begin{figure}
\centering
\subfloat{
\includegraphics[width=\columnwidth]{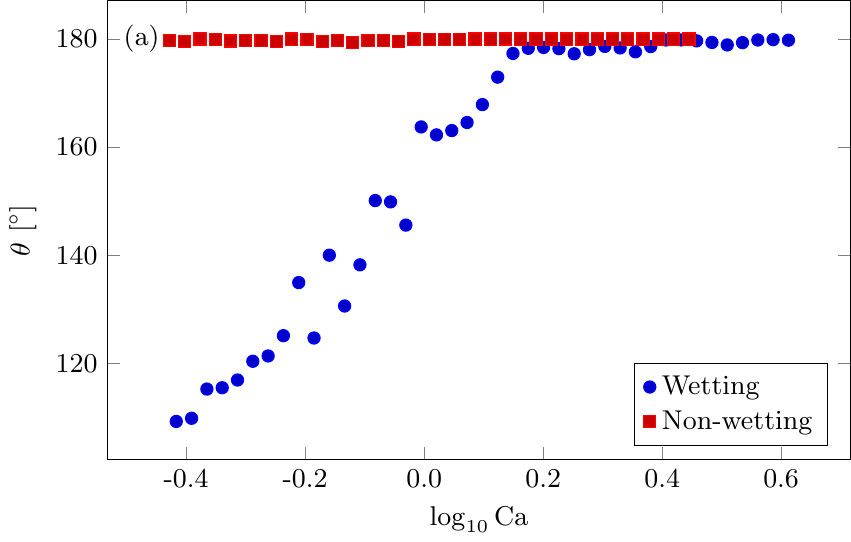}
} \\
\subfloat{
\includegraphics[width=\columnwidth]{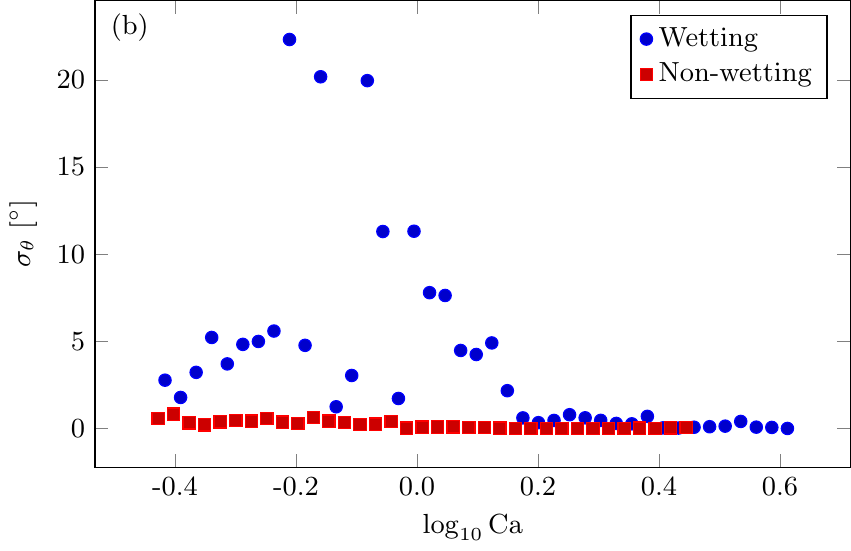}
}
\caption{Image (a) shows the average mean contact angle as a function of the capillary
number for a simulated droplet with the parameters of ethanol at atmospheric
pressure. For wetting surfaces a transition can be observed for $\txt{Ca}
\approx 1$ where the microscopic angle goes from $\theta < \ang{180}$ to $\theta
= \ang{180}$. For the non-wetting surface the microscopic angle stays constant.
Image (b) shows the standard deviation of the contact angle as a function of the
capillary number for a simulated droplet with the parameters of ethanol at
atmospheric pressure. For wetting surfaces, the contact line regime change
around $\txt{Ca} \approx 1$ causes the fluctuations of the microscopic contact
angle to increase significantly. This increase in fluctuations is not observed
for non-wetting surfaces.}
\label{fig:angleCa}
\end{figure}

A quantitative description of the differences in contact line behavior for
splashing on wetting versus non-wetting surfaces is provided in
Fig.~\ref{fig:angleCa} (a) by plotting the microscopic contact angle as function
of the capillary number for wetting and non-wetting surfaces. In agreement with
Fig.~\ref{if}, for $\txt{Ca} > 1$ both wetting and non-wetting surfaces show the
same contact angle of $\ang{180}$. When the contact line slows to $\txt{Ca} < 1$
the non-wetting surface continues to exhibit $\theta=\ang{180}$, while the
wetting surface exhibits a contact angle that decreases with $\txt{Ca}$. 

The change of behavior at $\txt{Ca} \approx 1$ is shown via the standard
deviation of the microscopic contact angle in Fig.~\ref{fig:angleCa} (b). For
the non-wetting surface $\theta=\ang{180}$ at all times, therefore the
fluctuations are small and independent of the capillary number. For the wetting
surface at large capillary numbers the fluctuations are also small, however
shortly before $\txt{Ca} = 1$ they begin to grow dramatically as $\txt{Ca} \to
1$. This behavior is reminiscent of a phase transition, where fluctuations
increase around the critical point. Taking this analogy further, this suggests
that not only does the contact angle change for $\txt{Ca} < 1$, but, more
importantly, the flow enters a different flow regime. Note that this result is
unrelated to the aforementioned touch-down events, which are instabilities of
the apparent contact angle. 

\begin{figure}
\centering
\subfloat{
\includegraphics[width=\columnwidth]{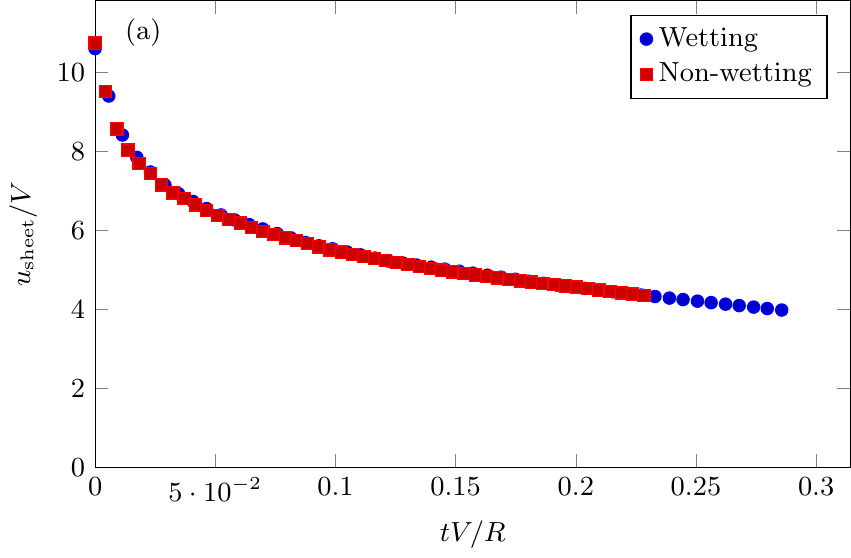}
} \\
\subfloat{
\includegraphics[width=\columnwidth]{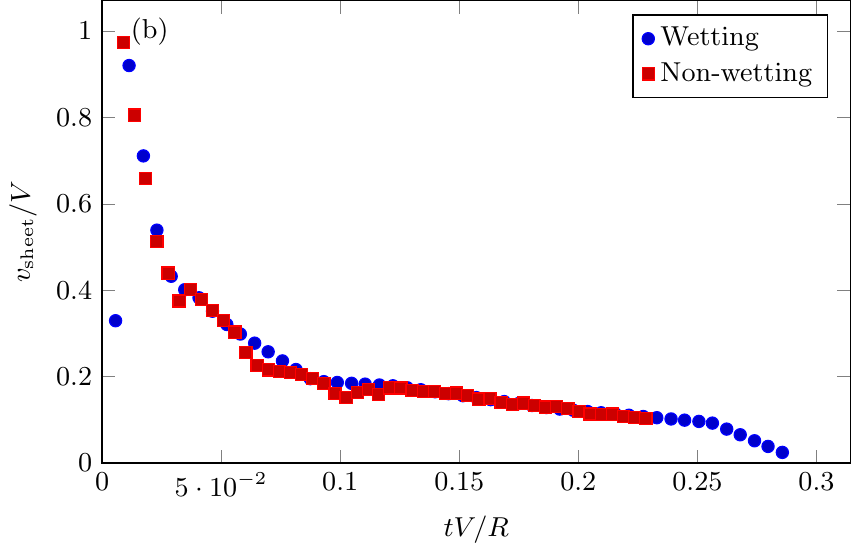}
}
\caption{Image (a) shows the horizontal liquid sheet ejection velocity as a
function of time for a simulated droplet with the parameters of ethanol at
atmospheric pressure. Velocities and time are made non-dimensional with the
impact velocity $V$ and droplet radius $R$. Image (b) shows the vertical liquid
sheet ejection velocity as a function of time for a simulated droplet with the
parameters of ethanol at atmospheric pressure. Velocities and time are made
non-dimensional with the impact velocity $V$ and droplet radius $R$. Wetting and
non-wetting surfaces show the same behavior.}
\label{fig:sheetTime} 
\end{figure}

Fig.~\ref{fig:sheetTime} shows the horizontal and vertical liquid sheet ejection
velocities, measured at the rim of the liquid sheet. Fig.~\ref{if} shows that a
cusp forms in the interface at the same time for both wetting and non-wetting
surfaces. Consequently, one can expect the liquid sheet, which forms promptly
after the cusp can be detected, to be ejected at the same velocity. This is
confirmed in Fig.~\ref{fig:sheetTime}, which demonstrates that the liquid sheets
are ejected at the same angle, independent of wetting properties. As time
progresses the figures show that the trajectories of the liquid sheet are nearly
identical for the wetting and non-wetting case.

%===================================================
\section{Discussion}
%===================================================
The key to understanding the unusual contact line behavior in splashing is in
contrasting it with that of slow-moving contact lines. A stationary contact line
will approach a homogeneous surface at an angle $\theta_{0}$ that is determined
purely by wetting properties. If the contact line is forced to move, as in the
classic case of a solid being plunged into a liquid bath, the
shape of the interface will be determined by both the capillary number, which
describes the balance between surface tension and viscous forces, and gravity
\cite{marchand2012}. In addition, the assumption is made that the contact line
is moving slowly enough that the microscopic contact angle is independent of the
capillary number and is equal to Young's angle $\theta_{0}$, which now serves as
a boundary condition at the surface. For advancing contact lines this results in
a critical capillary number at which this boundary condition cannot be satisfied
\cite{marchand2012,benkreira2008}. Consequently, theory predicts that above this
critical capillary number wetting failure will be observed in the form of air
bubble entrainment at the contact line. Additionally, it is predicted that the
critical capillary number depends on the wetting properties of the surface
\cite{riboux2014}.

The splash of the liquid drop occurs in multiple stages. Shortly before impact,
the bottom surface of the drop is deformed by the rising gas pressure in the
decreasing gap between the liquid and solid
\cite{mandre2009,mandre2012mechanism}. When the liquid makes contact with the
substrate, the air directly beneath the drop is trapped into a small bubble
confined to the center of the deposited liquid
\cite{chandra1991collision,Thoroddsen2010}, while the liquid begins spreading
radially outward in the form of an axisymmetric lamella
\cite{mongruel2009early}. Our simulations of lamella creation, described in more
detail in Ref.~\citenum{boelens2016a}, are consistent with the predictions made
in Refs.~\citenum{mandre2009},~\citenum{mongruel2009early},~and~\citenum{riboux2014}. The contact line moves fastest
immediately after impact, and proceeds to rapidly decelerate as shown in
Fig.~\ref{fig:uL}. Fig.~\ref{fig:uL} further reveals that at the moment of
thin-sheet formation the capillary number (on the right axis) of the contact line is in the unstable
regime: at atmospheric pressure $t_{\txt{sheet}}=0.21 \si{\m \s}$ with
$\txt{Ca}(t_\txt{sheet}) = 2.9$. Therefore at all times between impact and sheet
creation $\txt{Ca} > 2.9$.   

The time of thin-sheet creation, $t_{sheet}$, varies with multiple parameters, most
importantly with the ambient gas pressure. However, in all cases we find that
thin-sheet creation occurs when $\txt{Ca}(t_\txt{sheet}) \gtrsim 1$. Indeed, for
the points shown in Fig.~\ref{fig:tSheet} (a) a thin-sheet is created with $1.2 <
\txt{Ca}(t_\txt{sheet}) < 7.5$, for the 32 cP drops in Fig.~\ref{fig:tSheet} (b)
$1.6 < \txt{Ca} (t_\txt{sheet}) < 5.1$, and for the 9.4 cP drops in
Fig.~\ref{fig:tSheet} (b) $0.7 < \txt{Ca}(t_\txt{sheet}) < 3.0$. A wide range of
parameters was investigated in Ref.~\citenum{latka2016thin}. Impact velocity, drop size,
surface tension, density, surface tension, viscosity of both the liquid and the
gas, and the gas molecular weight were varied. It was invariably found that
splashing can occur only when the contact line is moving at a large Ca. 

The high resolution of our simulations allow us to determine that at such high
$\txt{Ca}$ the contact line is advancing via a "rolling" motion
\cite{boelens2016a}. In both the wetting, $\theta_0 = \ang{0}$, and non-wetting,
$\theta_0 = \ang{180}$, case, an ultra-thin air gap extends underneath the drop, as can
be seen in Fig.~\ref{if}(a-c). This is equivalent to a dynamic contact angle
$\theta=\ang{180}$ that is independent of the static wetting properties. In
other words, splashing is independent of wetting, because splashing occurs at
large $\txt{Ca}$, at which the static wetting properties do not influence the
shape of the contact line. Although we can identify this air film in
simulations, in experiments we are only able to detect the thicker gas film that
is present after $t_\txt{sheet}$. However, the ultra-thin air film was already
experimentally observed by Kolinski et al. via total internal reflection
measurements \cite{kolinski2012} and is consistent with the analysis in
Ref.~\citenum{rein2008}.

The presence of this air film can have a profound effect on drop impact. If the
substrate is sufficiently smooth and the impact velocity sufficiently low, the
ultra-thin air persists underneath the drop and isolates it from substrate, so
that a drop can rebound from a wetting substrate, as if it were
super-hydrophobic \cite{kolinski2014drops}. However, drops splash at much higher
impact velocities. In this case, Kolinski et al. observed that the air film
behind the advancing contact line was closed within several microseconds, and
attributed the effect to interactions between the liquid and solid
\cite{kolinski2012,wyart1990drying,yiantsios1991rupture}. This is consistent
with the observation that while splashing, which originates directly at the
contact line, is independent of the wetting properties for both contact angles
smaller than $\ang{90}$ and larger than $\ang{90}$ \cite{goede2017}, the slower dynamics
following splashing are still determined by the static contact angle. For
example, the maximum spreading diameter of an impacting drop does depend on the
properties of the solid \cite{wildeman2016on,antonini2012drop,vsikalo2005dynamic}. 
We emphasize that the
wetting independence of splashing is a direct result of contact line dynamics.
While our simulations agree with Ref.~\citenum{philippi2016drop} with respect to the
presence of a boundary layer and also find a characteristic "rolling" motion of
the contact line, we find, in agreement with Ref.~\citenum{kolinski2012}, that these
phenomena result from the rapid motion of the contact line.

The behavior of the splashing contact line at high Ca provides a means of
testing splashing theories, such as the model recently proposed by Riboux and
Gordillo \cite{riboux2014,riboux2015} or by Liu et al. \cite{liu2015}. In the
Riboux and Gordillo model the wetting properties influence the calculated
lubrication lift force on the edge of the spreading liquid. As the liquid edge
rises, its rim increases in size due to surface tension and, consequently, the
bottom surface of the rim is forced downward. Depending on which of these
effects dominates, the lamella either continues moving upward and eventually
breaks apart to form a splash, or rewets the substrate, which prevents
splashing. The lubrication force calculated is dependent strongly on the shape
of the advancing contact line set by the microscopic contact angle, (c.f.
footnote [35] in Ref.~\citenum{riboux2014}). Consequently, the theory predicts the
splashing threshold to strongly depend on the wetting properties of the
substrate; this is inconsistent with the results presented here from experiments
and simulations. Furthermore, this model predicts that splashing occurs at time
$t_{e,crit}$ after impact, calculated from  Eqn. 1 in Ref.~\citenum{riboux2014}. In
contrast, we find that in all cases $t_{\txt{sheet}}\gg t_{e,crit}$
\cite{driscoll2010,driscoll2011,stevens2014b}. The disparity is most evident for
drops of higher viscosity near threshold pressure, for which $t_{\txt{sheet}}$
is largest. For these drops, $t_{e,crit}$, which does not depend on pressure, is
smaller than $t_\txt{sheet}$ by several orders of magnitude. Together, these
results suggest that the microscopic basis of this and similar theories should
be revisited. In contrast, Liu et al. propose that splashing is caused by the
Kelvin-Helmholtz instability in the air film that was observed in
Ref.~\citenum{kolinski2012} and in our simulations. While our results are not a
direct test of this model, they are consistent with its implicit contact angle
independence.  

%===================================================
\section{Conclusion}
%===================================================
Splashing arises from the interaction of three phases: the liquid drop, the
solid substrate, and the ambient gas. It is therefore surprising that the most
basic measure of this interaction, the contact angle $\theta_{0}$, does not
influence the outcome of drop impact for the impact parameters we have
investigated. Our experiments show that both the time of the splash, as well as
the splashing thresholds are independent of $\theta_{0}$. Direct numerical
simulations allow us to probe the advancing liquid-gas interface at nm length
scales and show that the shape of this interface is the same for $\theta_{0} =
\ang{0}$ and $\theta_{0} = \ang{180}$. 

Splashing occurs when the liquid is spreading rapidly across the substrate, at
capillary numbers $\txt{Ca} \ge 1$. In this regime, both experiments
\cite{kolinski2012} and simulations \cite{boelens2016a} suggest that the
advancing contact line spreads over a short-lived thin film of air.
Understanding the dynamics of this air film, both its rapid growth at
$t_{\txt{sheet}}$ (where we define $t_{sheet}$ as the time when the thin sheet
first starts to appear at the front of the expanding lamella) that leads to
splashing, and how $t_{\txt{sheet}}$ is set by ambient pressure, is crucial to
forming an accurate model of splashing.

%===================================================
\begin{acknowledgments}
This work was supported by the University of Chicago Materials Research and
Engineering Center (MRSEC) through grant DMR-1420709 and by the NSF grant
DMR-1404841. The VOF simulation method for splashing employed here was developed
with support from the Multi-University Research Initiative (MURI) program -
Office of Naval Research (MURI) - N00014-11-1-0690.  
\end{acknowledgments}
%===================================================

%===================================================
%\bibliography{article.bib}
%
%===================================================

\end{document}